\documentclass[aps,prl,twocolumn,superscriptaddress,longbibliography]{revtex4-1}
\usepackage[utf8]{inputenc}
\usepackage{graphicx}
\DeclareGraphicsExtensions{.png,.jpg,.eps,.pdf}

\usepackage{float}
\usepackage{xcolor}
\usepackage{amsmath}
\usepackage{float}
\usepackage[colorlinks=true,citecolor=blue]{hyperref}

\DeclareMathAlphabet\mathbfcal{OMS}{cmsy}{b}{n}

\begin{document}

\title{An ab initio description of the family of Cr selenides: structure, magnetism and electronic structure from bulk to the single-layer limit}

\author{Jan Phillips}
  \email{j.phillips@usc.es}
\affiliation{Departamento de F\'{i}sica Aplicada,
  Universidade de Santiago de Compostela, E-15782 Campus Sur s/n,
  Santiago de Compostela, Spain}
\affiliation{Instituto de Materiais iMATUS, Universidade de Santiago de Compostela, E-15782 Campus Sur s/n, Santiago de Compostela, Spain}  
\author{Adolfo O. Fumega}
\affiliation{Department of Applied Physics, Aalto University, 02150 Espoo, Finland}

\author{S. Blanco-Canosa}
\affiliation{Donostia International Physics Center (DIPC), San Sebastián, Spain}
\affiliation{IKERBASQUE, Basque Foundation for Science, 48013 Bilbao, Spain}

\author{Victor Pardo}
  \email{victor.pardo@usc.es}
\affiliation{Departamento de F\'{i}sica Aplicada,
  Universidade de Santiago de Compostela, E-15782 Campus Sur s/n,
  Santiago de Compostela, Spain}
\affiliation{Instituto de Materiais iMATUS, Universidade de Santiago de Compostela, E-15782 Campus Sur s/n, Santiago de Compostela, Spain}

\begin{abstract}

Compounds based on Cr have been found to be among the first single-layer magnets. In addition, transition metal dichalcogenides are promising candidates to show long-range ferromagnetic order down to the two-dimensional limit. We use ab initio calculations to provide a description of the various Cr$_x$Se$_{x+1}$ stoichiometries that may occur by analyzing from the bulk materials to the monolayer limit. We study the different structural distortions, including charge density waves that each system can present by analyzing their phonon spectra and dynamic stability. We provide a description of their basic electronic structure and study their magnetic properties, including the magnetocrystalline anisotropy energy. The evolution of all these properties with the dimensionality of the systems is discussed. This intends to be a comprehensive view of the broad family of Cr selenides.

\end{abstract}


\maketitle

\section{Introduction}\label{sec:intro}
Transition metal dichalcogenides (TMD) have become a family of materials under intense study in recent years. Their laminar structure has allowed to create very thin layers of them, down to the monolayer limit, where the community has explored all sorts of interesting physical properties: magnetism\cite{sethulakshmi2019magnetism}, topological properties\cite{li2017evidence}, charge density waves\cite{xi2015strongly,rossnagel2011origin,neto2001charge}, superconductivity\cite{yokoya2001fermi,sipos2008mott}, heavy fermions\cite{vavno2021artificial}, etc. They have become central in the broader world of two-dimensional (2D) materials as key pieces to form heterostructures\cite{novoselov20162d,manzeli20172d}: artificial materials formed by stacking different units, weakly bonded by van der Waals forces, that as a whole present properties that do not exist in the different layers separately. Recently, the additional degree of freedom of the twisting angle to form moiré patterns has been incorporated in the playground\cite{angeli2021gamma,rosenberger2020twist}, that certainly does not exist when dealing with bulk materials. All this makes TMD's an exceptional framework to explore new applications, test theories, and analyze all sorts of phenomena that emerge in Condensed Matter Physics.

Among them, Cr-based van der Waals compounds are currently under intensive research motivated by the observation of ferromagnetism at the atomic limit, such as in CrI$_3$ \cite{huang2017layer, gong2017discovery, Lado_2017}, Cr$_2$Ge$_2$Te$_6$ \cite{Xing_2017}, or CrSBr \cite{Ziebel2024}. More recently, kagome-like bands have been observed in the Cr$_8$Se$_{12}$ (Cr$_2$Se$_3$) system \cite{duan2024observation}, opening the path to engineering strongly correlated frustrated lattices and the emergence of novel magnetic and electronic phases of matter. 

Cr chalcogenides with quasi-2D structure can exist in multiple forms and stoichiometries. Different phases even with non-uniform stoichiometries may exist and compete in the same sample in the form of nanoflakes that grow with different properties, which makes it extremely complicated to characterize the material at hand. The goal of our work is to provide a comprehensive study of the broad range of Cr-based selenide materials (Cr$_x$Se$_{x+1}$) based on ab initio calculations and to identify their basic characteristics from the structural, electronic and magnetic point of view.

We will analyze a variety of Cr-based compounds, engineering structures based on the experimental reports \cite{doi:10.1021/acs.jpclett.1c01493, 1973crse, YUZURI1977891, sato1990reflectivity, YUZURI1987223, OHTA1997168, Adachi1994, https://doi.org/10.1002/adfm.201805880, doi:10.1021/acs.nanolett.9b00386, doi:10.1021/acs.chemmater.1c01222, PhysRevB.87.014420, van1980crse2, PhysRevB.87.014420, bruggen_crse2_1980, MYRON1981120, PhysRevB.89.054413, doi:10.7566/JPSJ.84.064708, luout}, studying the effect of their structural dimensionality, stoichiometry and strain.  The manuscript is organized as follows. In Section \ref{sec:comp}, we describe the computational details. In Section \ref{sec:struct} we present the structural analysis of the different Cr$_x$Se$_{x+1}$ stoichiometries and polymorphs, including CrSe$_2$, a Cr-based van der Waals material. We begin with the 1\textit{T} phase, and find that the high symmetry phase is a magnetic conductor that is not dynamically stable from the bulk limit, down to the monolayer. In Section \ref{sec:ES}, we will focus on the electronic structure and magnetsim of the different compounds. Section \ref{sec:cdw} will delve into the study of phonon instabilities and possible charge density waves, and in Section \ref{sec:MAE}, we will explore the emergence of long-range magnetic order in the low-dimensional limit. Finally, our conclusions are summarized in Section \ref{sec:conclusions}.    


\section{Computational methods}\label{sec:comp}

We have performed \emph{ab initio} electronic structure calculations based on the density functional theory (DFT)\cite{HK,KS} using an all-electron full potential code ({\sc wien2k}\cite{WIEN2k}).
The exchange-correlation term used for all structures was the generalized gradient approximation (GGA) in the Perdew-Burke-Ernzerhof\cite{PBE} (PBE) scheme. We also used the Vienna Ab Initio Package ({\sc VASP}) code \cite{kresse1993ab,kresse1996efficiency,kresse1996efficient}, where again we used GGA-PBE based pseudopotentials. Note that in the case of TMD's, van der Waals schemes may be required to analyze structural properties\cite{diego2021van}.  However, our structural analysis that focused on monolayer limits did not need any additional schemes because of the absence of this van der Waals gap. When treating with bulk structures with van der Waals gaps, we relaxed the structures with {\sc VASP} using the DFT-D3 method with Becke-Johnson damping function \cite{grimme_vdW}.
All calculations were performed with a converged k-mesh using the Monkhorst-Pack scheme\cite{monkhorst_pack}, and using the relaxed lattice parameters and internal positions of the structures. We did not find a significant difference between the {\sc wien2k} code and the {\sc VASP} code in the structural parameters after the relaxations. All the electronic structure results where obtained using {\sc wien2k}. The harmonic phonon spectrum of all compounds was computed using the real-space supercell approach\cite{phonopy, phonopy-phono3py-JPSJ}, with the structures obtained from {\sc VASP} relaxations.


\begin{figure*}
  \includegraphics[width=\textwidth]%
    {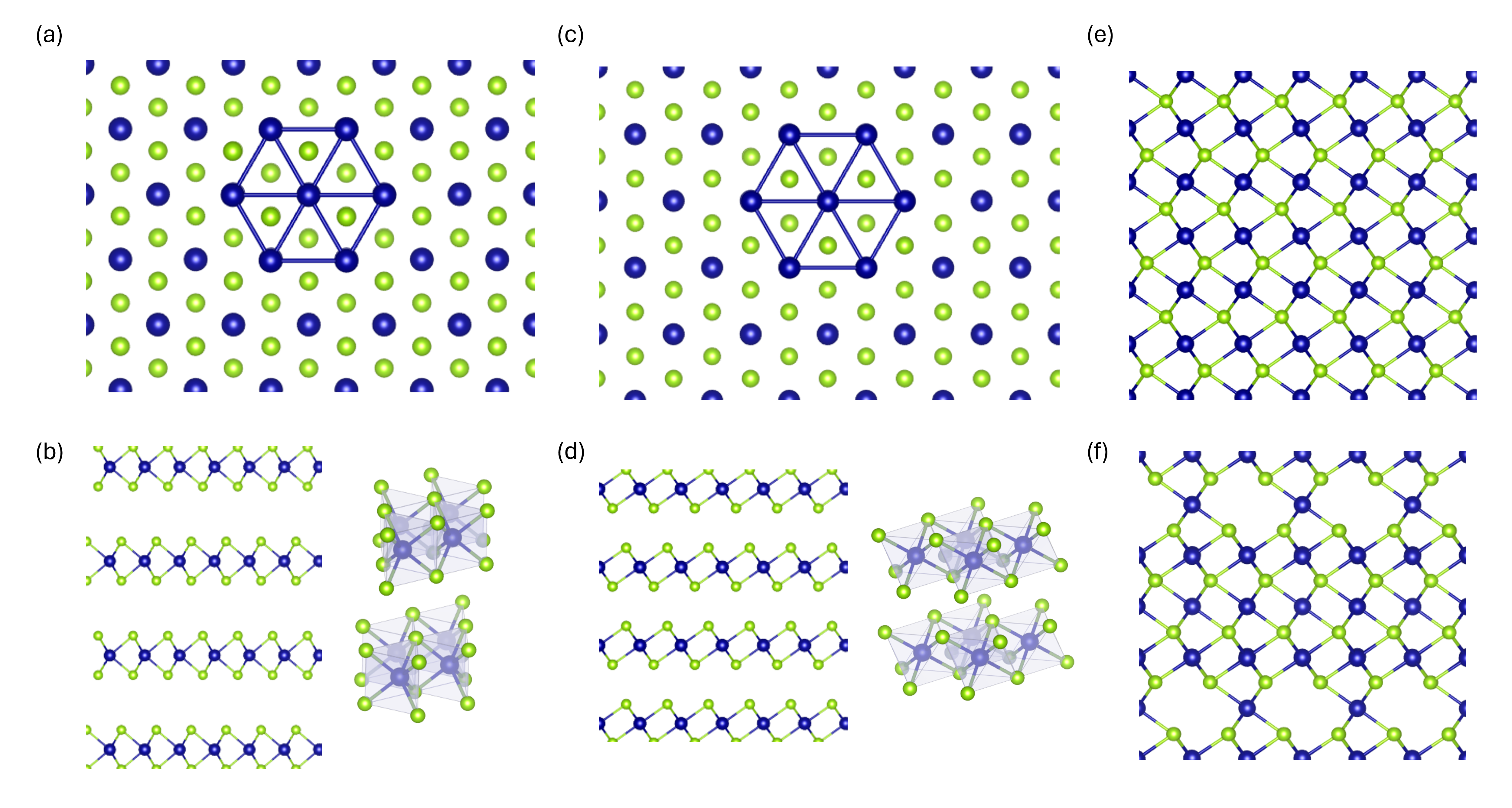}
     \caption{Cr$_x$Se$_{x+1}$ 2-\textit{H} structure from (a) a top view and (b) a side and slanted view, where the trigonal environment can be seen.  Cr$_x$Se$_{x+1}$ 1-\textit{T} structure from (c) a top view and (d) a side and slanted view, where the octahedral environment can be seen. NiAs-type structure from a side view with no Cr vacancies (e) and with some Cr vacancies (f).}\label{struct}
\end{figure*}

In order to compute the magnetic anisotropy energy (MAE), spin-orbit coupling (SOC) was introduced in a second variational manner using the scalar relativistic approximation \cite{SOC_Macdonald}. MAE is defined as:
\begin{equation}\label{eq_MAE}
    MAE=E_{in}-E_{out},
\end{equation}
where $E_{in}$ is the energy per metal atom when the magnetization is set along an in-plane direction (parallel to the Cr layers), while $E_{out}$ is calculated considering the magnetization points along the out-of-plane direction (perpendicular to the Cr-rich layers). An out-of-plane MAE (positive values in our convention) when working with the purely 2D monolayers (i.e. in the case of CrSe$_2$) circumvents the restrictions of the Mermin-Wagner theorem\cite{mermin1966absence} leading to the possibility of long-range ferromagnetic (FM) order to be stable at finite temperatures.

\section{Structural Analysis}\label{sec:struct}

In this work, we present ab initio calculations for different structures and stoichiometries of Cr$_x$Se$_{x+1}$. This system offers a rich variety of structural possibilities and can grow either as a 2D van der Waals crystal or with a 3D Volmer-Weber kind of growth \cite{doi:10.1021/acs.jpclett.1c01493}. Because of this, characterizing these structures at an atomic scale using techniques such as scanning tunnel microscopy (STM) can be a complicated task. Most of these stoichiometries present similar growths and surfaces which do not give enough information about other structural details such as van der Waals gaps or the presence of vacancies in the system. We study a wide range of stoichiometries, analyzing how the different structures and environments affect the electronic and magnetic properties of each system. Typical TMDs present van der Waals structures in their high symmetry phases. These are commonly either a 1\textit{T}, as in Fig. \ref{struct}(c-d) where the chalcogen atoms form an octachedral coordination surrounding the transition metal atom, or 2\textit{H}, as in Fig. \ref{struct}(a-b) where the chalcogen coordination is trigonal. We refer to these high symmetry phases as normal states (NS), in opposition to the distorted states that appear due to, e.g.,  charge density waves (CDW). From a perpendicular view to the Cr layers, both these high symmetry phases are hard to tell apart, as can be seen by comparing Fig. \ref{struct}(a) and (c). This is a difficulty which is present when characterizing this kind of system up to an atomic scale using techniques like STM. A side view of both of these phases is depicted in Fig. \ref{struct}(b) and (d), where it can be seen that both present a van der Waals gap. 
The Cr$_x$Se$_{x+1}$ system can grow without a van der Waals gap through a Volmer-Weber 3D crystal growth, in a \textit{T} symmetry type of phase, like the one shown in Fig. \ref{struct}(e), which can vary its stoichiometry depending on its height or number of Cr-layers. From the analysis of this structure, which presents no vacancies, we observe a relation between the number of Cr layers and the stoichiometry of the form Cr$_L$Se$_{L+1}$, where L is the number of Cr layers per unit cell. Therefore, Cr-rich layers present a NiAs-type structure. 

Playing around with the number of layers allows to computationally analyze different stoichiometries that serve as a minimal unit to possible van der Waals layered heterostructures. We will analyze the case for 3 and 4 layers and CrSe as the infinite layer limit of a NiAs-type structure. This kind of structure commonly presents Cr vacancies, as depicted in Fig. \ref{struct}(f). We will comment on this in more detail when discussing the Cr$_2$Se$_3$ stoichiometry below.  We will also pay special attention to the particular case of the 2D limit of the NiAs structure, with only one Cr layer, which corresponds to monolayer 1\textit{T}-CrSe$_2$. We present results for this stoichiometry from bulk to monolayer for both the \textit{T} and \textit{H} phases.  Our focus is always in approaching the 2D limit, where we pay special attention to possible ground states and their magnetism.

\subsection*{The bulk limit}
The most stable form of the Cr-Se series is Cr$_2$Se$_3$, most other stoichiometries decomposing to it in some part of their phase diagrams. Bulk Cr$_2$Se$_3$ is known to have a NiAs (space group $P6_3/mmc$) kind of structure with ordered Cr-ion vacancies, with a nominal Cr$^{3+}$ valence\cite{1973crse}. It is an antiferromagnetic (AF) semiconductor with a Néel temperature around 43 K \cite{YUZURI1977891, sato1990reflectivity}. The magnetic properties can change by applying pressure or through chemical substitution \cite{YUZURI1987223, OHTA1997168}. The experimental in-plane lattice parameters are $a=b=6.25$ \r{A} and $c=17.27$ \r{A} \cite{Adachi1994}. One can picture the Cr$_2$Se$_3$ structure as Cr layers using a unit cell consisting of a $\sqrt{3}\times\sqrt{3}$ supercell of the 1\textit{T} high symmetry (normal state) unit depicted in Fig. \ref{struct}(a), with each layer containing 3 Cr atoms in the unit cell. The bulk structure presents alternating Cr-rich layers and layers with Cr vacancies, following a similar pattern to the structure represented in Fig. \ref{struct}(f). Essentially, Cr$_2$Se$_3$ has layers that present two Cr vacancies in between Cr-rich layers. The Se atoms in this structure form an octahedral environment surrounding the Cr atoms, see Fig. \ref{struct}(d), without van der Waals gap. Thus, exfoliating Cr$_2$Se$_3$ could result in different Cr$_x$Se$_{x+1}$ stoichiometries. Monoflakes of Cr$_{15}$Se$_{18}$, i.e. (Cr$_{2.5}$Se$_{3}$ stoichiometry) consisting on 3 Cr rich layers and 2 Cr-vacancy layers, have been reported in the literature \cite{https://doi.org/10.1002/adfm.201805880, doi:10.1021/acs.nanolett.9b00386}. 
Moreover, Cr$_{2}$Se$_{3}$ nanoflakes down to 1.9 nm thickness have been grown via ambient pressure chemical vapor deposition \cite{doi:10.1021/acs.chemmater.1c01222}, that would correspond to approximately one bulk unit cell. According to our phonon calculations, presented in Fig. S1(d), we see that the bulk structure of Cr$_2$Se$_3$ is stable with relaxed in-plane and out-of-plane lattice parameters of $a=6.32 $ \r{A} and $c=17.50$ \r{A}, respectively. Due to the fact that the in-plane structure of the unit cell of bulk Cr$_2$Se$_3$ is equivalent to a $\sqrt{3}\times\sqrt{3}$ reconstruction of the NS unit cell, if unaware of the internal Cr vacancies in the structure, the bulk unit cell could easily be mistaken by a $1\times1$ NS unit cell with an in-plane lattice parameter of \textit{a}=3.65 \r{A}. 

As mentioned previously, it is not possible to get closer to the 2D limit and maintain both the octahedral environment and the stoichiometry of Cr$_2$Se$_3$. By reducing this structure to a single Cr-rich monolayer, the stoichiometry would change to 1\textit{T}-CrSe$_2$, which in itself has a very different set of physical properties compared to Cr$_2$Se$_3$. Antiferomagnetic CrSe$_2$, T$_N$=48  K \cite{CMFang_1997,PhysRevB.87.014420,van1980crse2,PhysRevB.87.014420}, is known to be a metastable form of the Cr$_x$Se$_{x+1}$ series \cite{bruggen_crse2_1980,MYRON1981120} and it has been reported to present different phase transitions.\cite{PhysRevB.89.054413, doi:10.7566/JPSJ.84.064708}.
The calculations in Ref. \cite{PhysRevB.87.014420} show that the ground state couplings are FM within the plane and AFM between planes, with other AFM phases competing in energy as a function of strain and doping. Therefore, the competition between FM and various AFM couplings can be removed by going down to the monolayer limit, where only the in-plane FM interaction would survive. The structure of bulk CrSe$_2$ in the 1\textit{T} phase can be seen in Fig. \ref{struct}(c-d). The unit cell is triangular, with the Cr atoms forming a hexagonal lattice with relaxed in-plane and out-of plane lattice parameters of $a=3.50$ \r{A} and $c=5.75$ \r{A}, respectively. From our phonon band structure calculations (shown in Fig. S5(d)), we see that bulk  1\textit{T}-CrSe$_2$ is not dynamically stable. We will discuss the reconstruction of the CrSe$_2$ stoichiometry for the monolayer case in Section \ref{sec:cdw}. First principles calculations have predicted that the \textit{H} phase is the lower-energy structure in the monolayer CrX$_2$ series\cite{doi:10.1021/jp501734s} and it can be connected to the \textit{T}-phase through strain\cite{doi:10.1021/jp501734s}. The bulk structure presents a 2\textit{H}-phase structure, Fig. \ref{struct}(b), where the main difference with the \textit{T}-phase is the trigonal environment around the Cr atoms (octahedral in the \textit{T}-phase). The relaxed in-plane and out-of-plane lattice parameters of the bulk structure are $a=3.20$ \r{A} and $c=13.56$ \r{A}, respectively. Our phonon band structures (shown in Fig. S10(d)) present no imaginary modes, in clear contrast with the unstable  1\textit{T} bulk structure, where the high-symmetry phase is not dynamically stable.

A specific case of interest that must be addressed is the structure of the Cr$_x$Se$_{x+1}$ family without van der Waals gap between Cr layers and vacancy layers. This case would correspond, in the bulk structure, to CrSe with a NiAs-type configuration and rhombohedral symmetry (space group \textit{R}$\bar{3}$). It consists of Cr layers with an octahedral Se environment, as shown in Fig. \ref{struct}(e). A single layer of this Cr-full NiAs-type structure coincides with a CrSe$_2$ \textit{T}-phase monolayer. Stacking another Cr-Se layer on top of this would lead to two NiAs-type layers in such a way that the overall stoichiometry is Cr$_2$Se$_3$. Adding layers in this same manner allows us to study and engineer Cr$_L$Se$_{L+1}$ stoichiometries which may serve as minimal units to form new heterostructures with a van der Waals gap. The infinite-layer system leads us back to the bulk CrSe system, as in Fig. \ref{struct}(e), with a relaxed in-plane lattice parameter $a=3.76$ \r{A}. This configuration presents a dynamically stable phonon band structure (Fig. S15(d-e)).

\subsection*{Approaching the 2D limit}

\begin{figure}
  \includegraphics[width=0.5\textwidth]%
    {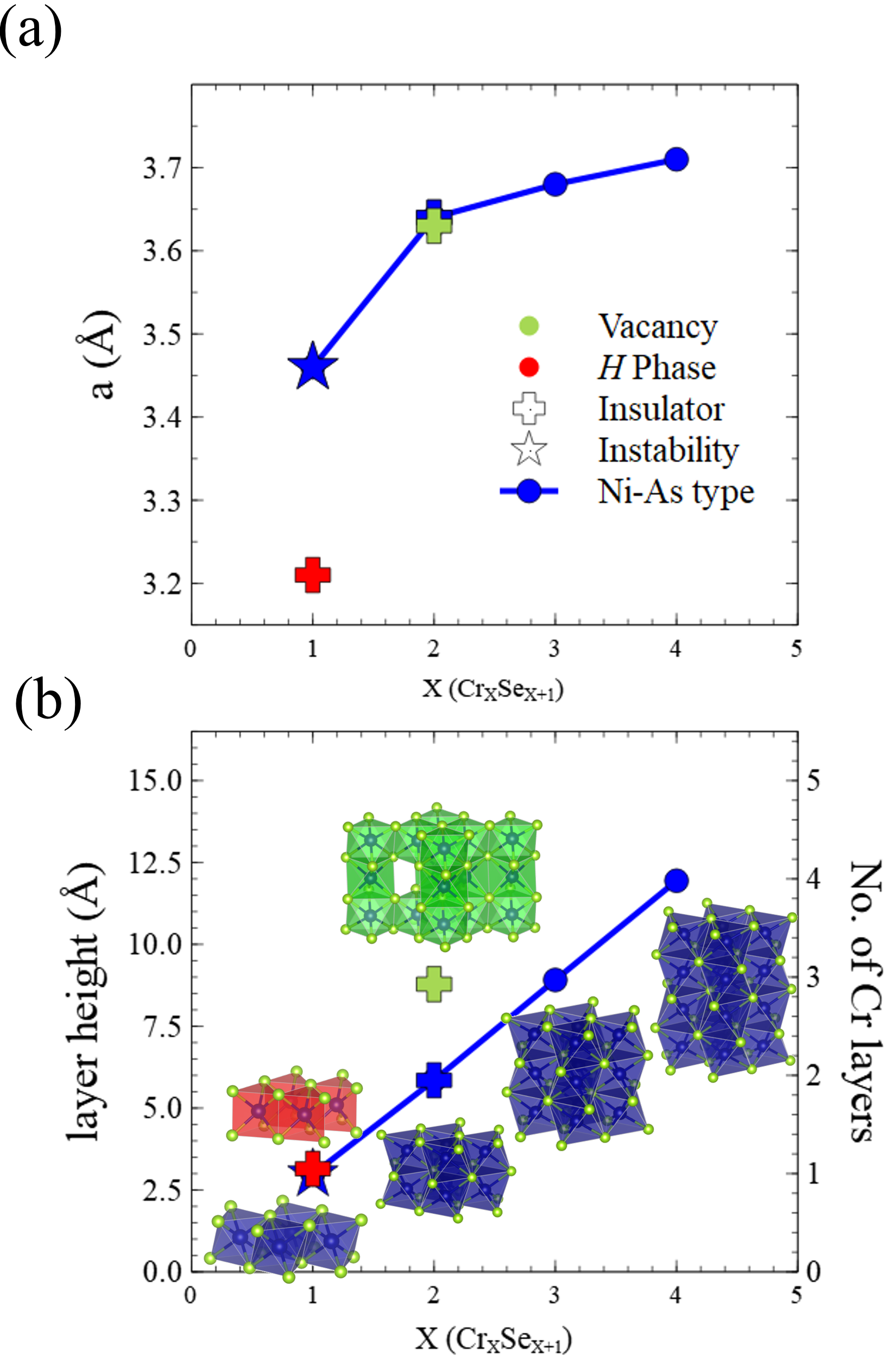}
     \caption{Cr$_x$Se$_{x+1}$ van der Waals units. (a) In-plane lattice parameter as a fucntion of Cr ($x$) and Se ($x+1$) in the system (stoichiometry). (b) Layer height (number of layers) as a function of Cr ($x$) and Se ($x+1$) in the system (stoichiometry).}\label{vdw_units}
\end{figure}

In Fig. \ref{vdw_units}, we present results of the different van der Waals minimal units we have analyzed. From Fig. \ref{vdw_units}(b) wee see that the height of the minimal van der Waals units does not depend on the stoichiometry, but on the number of Cr layers. This differs from Fig. \ref{vdw_units}(a), where the in-plane lattice parameter increases with the amount of Cr. We observe a linear decrease of the layer height in all of the NiAs-type \textit{T}-phase structures when approaching the 2D limit of CrSe$_2$, which presents a similar height for both the \textit{T} and \textit{H} phases. Analyzing the in-plane lattice parameter, the decrease is no longer linear down to the 2D limit. In this limit, a change in the structural symmetry of the system also affects the in-plane lattice parameter. This can be seen in Fig. \ref{vdw_units}(a) by comparing the two different phases for CrSe$_2$. The different coordination of the Se atoms surrounding Cr greatly reduces the in-plane lattice parameter. In general, all the higher-symmetry phases are dynamically stable, with the exception of CrSe$_2$ in its 1\textit{T}-phase that presents instabilities in its phonon band structure, inherited from the bulk structure down to the monolayer. We will discuss this in further sections.

Starting with the case of Cr$_2$Se$_3$, the aforementioned complexity of its bulk structure makes it difficult to preserve its particular vacancy structure and access a lower-dimensional structure. In order to reduce the dimensionality, we have engineered an alternative structure consisting of a three-layer system that preserves the stoichiometry. This means that the Cr-vacancy layer has only one vacant Cr site. This leaves us with the Cr-vacancy layer sandwiched between two Cr-rich layers. This structure can be seen in Fig. \ref{vdw_units}(b). It comes as a natural approach towards the 2D limit, where the vacancy scheme is still present. The relaxed in-plane lattice parameter obtained from DFT in this case is $a=6.29$ \r{A}, very close to the bulk value and $a=3.63$ \r{A} in comparison with the typical $1\times1$ NS unit cell of TMDs. The computed phonon band structures of this monolayer (Fig. S2(d)) does not present imaginary frequencies, indicating the structure is stable under possible in-plane atomic reconstructions. 

We also propose another alternative monolayer which preserves the Cr$_2$Se$_3$ stoichiometry. Keeping an analogous approach to the previous structure, we removed the last vacancy left in the structure introducing another Cr atom, but without preserving the 2:3 ratio between Cr and Se atoms. Indeed, the stoichiometry can be preserved if the system consisted of 2 Cr layers, thus reducing our unit cell to only two Cr layers of a NiAs-type structure as shown in Fig. \ref{vdw_units}(b) (and S3(a-c)). The DFT structural relaxation returns in-plane lattice parameters comparable to those of typical high-symmetry $1\times1$ TMD's, $a=3.64$ \r{A}, very close to those provided for the previous monolayer (as shown in Fig. \ref{struct}(a)) and the bulk structure. The newly constructed Cr$_2$Se$_3$ monolayer also presents a stable phonon band structure, Fig. S3(d). 

Considering the last described monolayer, where the van der Waals unit consists of two Cr layers, as the minimal possible Cr$_2$Se$_3$ unit, we built a bilayer and evaluated the behavior of that hypothetical structure when approaching the bulk limit. This computationally engineered bilayer is dynamically stable (Fig. S4(d)) thus offering the possibility of a Cr$_2$Se$_3$ van der Waals structure which could be exfoliated down to the monolayer limit from its bulk structure.

Focusing now on 1\textit{T}-CrSe$_2$, we move forward by studying a bilayer, i.e. two van der Waals bonded layers of CrSe$_2$ in its 1\textit{T} phase, like described in Fig. \ref{struct}(d). The structure presents a relaxed in-plane lattice parameter $a=3.49$ \r{A}, similar to the bulk value. A careful inspection of the imaginary modes in the phonon dispersions (Fig. S6(d)) confirms that the main in-plane instabilities that appear in the bulk system are also present in the bilayer system. In the case of the 2\textit{H}-CrSe$_2$ bilayer, just like the previous structure, the bilayer shows no difference compared to the bulk structure. The phonon band structure shows no imaginary modes (Fig. S11(d)), with a very similar in-plane lattice parameter of $a=3.21$ \r{A}.

Due to the van der Waals bonded layers, the stoichiometry and structural details are easily preserved moving towards the 2D limit. We performed DFT calculations in the monolayer system of CrSe$_2$ in its 1\textit{T} phase. Its structure comes as a single Cr layer of the bulk structure (that can be seen in Fig. \ref{struct}(c-d)). The relaxed in-plane lattice parameter changes slightly in this case in comparison to the bulk and bilayer discussed above. DFT results yield a value of $a=3.46$ \r{A}. The phonon band structure presents instabilites (Fig. S7(d)) in a very similar way to that of the bulk and bilayer CrSe$_2$.

 In monolayer \textit{H}-CrSe$_2$, we obtain a relaxed in-plane lattice parameter $a=3.21$ \r{A}, identical to the bilayer case. The phonon band structure does not present any imaginary phonon modes (Fig. S12(d)), thus confirming our expectations of dynamic stability down to the monolayer limit. From inspection of Fig. \ref{vdw_units}(a), the in-plane lattice parameters for the two different symmetries analyzed for CrSe$_2$, \textit{T} and \textit{H},  are noticeably different, unlike their height, which is practically identical as seen in Fig. \ref{vdw_units}(b).

We finish with this Section analyzing the case of CrSe on approaching the 2D limit. Although the absence of a van der Waals gap in CrSe prevents the reduction of the structural dimensionality while preserving the stoichiometry, there is still interest in studying this system when approaching the 2D limit. CrSe is the infinite-layer limit of the NiAs-type structure, hence the 3- and 4-layered system corresponds to Cr$_3$Se$_4$ and Cr$_4$Se$_5$, respectively. Similar to Cr$_2$Se$_3$, the \textit{L}=3 and 4-layer cases are dynamically stable (Fig. S13(d-e) and Fig. S14(d-e)) and the relaxed lattice parameters of these stoichiometries are very close to each other, $a=3.68$ \r{A}, and $a=3.71$ \r{A}, for Cr$_3$Se$_4$ and Cr$_4$Se$_5$, respectively. This gives us two different stoichiometries that can work as van der Waals minimal units with a NiAs-type structure.

\section{Electronic structure and Mangetism}\label{sec:ES}
\begin{figure*}
  \includegraphics[width=\textwidth ]%
    {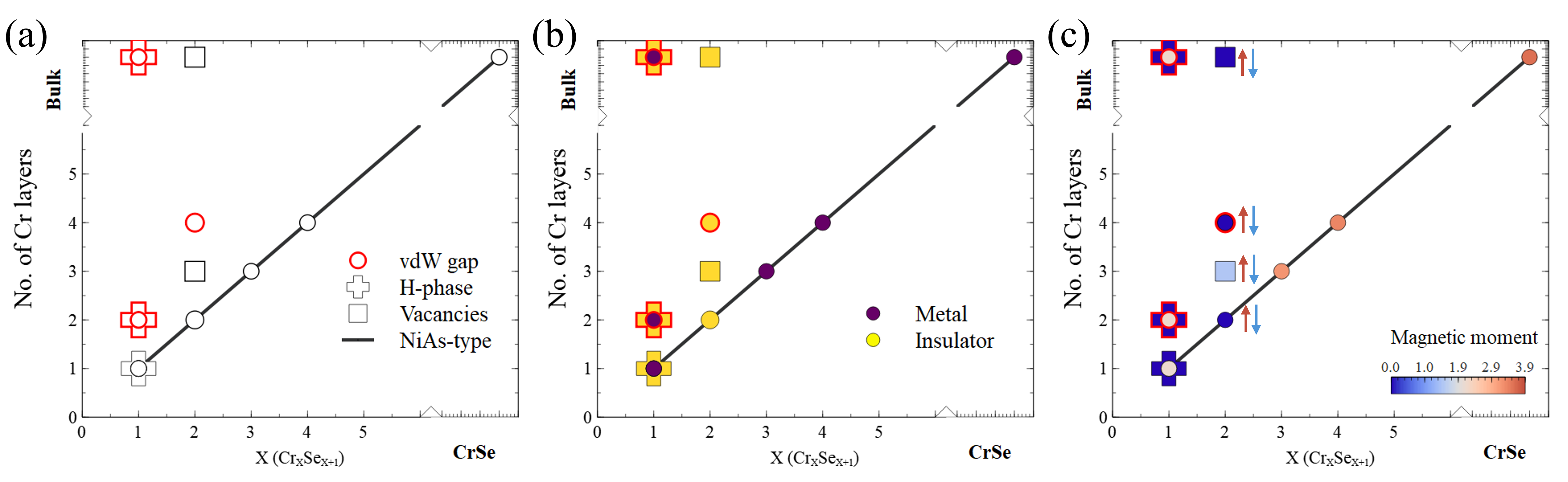}
     \caption{Electronic structure and magnetism of the Cr$_x$Se$_{x+1}$ family represented through the number of Cr layers as a function of the amount of Se (X). (a) Structural details of the analyzed structures from bulk to monolayer. Red marker borders indicate there is a van der Waals gap in the system, crosses indicate an \textit{H}-phase, squares indicate there are vacancies present in the structure and the black line indicates a NiAs-type structure as we increase stoichiometry (layers). (b) Metallic and insulating character where data marked in purple (yellow) indicates a metal (insulator). (c) The magnetic moment is represented in a color map, where blue (red) represents the minimum (maximum) computed value. Also, pairs of red and blue arrows indicate antiferromagnetic magnetic order in the system. }\label{ES_mag}
\end{figure*}

We have condensed our results of the electronic structure and magnetism in Fig. \ref{ES_mag}. All structures discussed previously have been analyzed and are represented here according to the references depicted in Fig. \ref{ES_mag}(a). We observe that the insulating character corresponds to certain stoichiometries, as is the case of the Cr$_2$Se$_3$ structures, and a certain Se environment, as in the \textit{H}-CrSe$_2$ layered family. This is shown in Fig. \ref{ES_mag}(b). The Cr$_2$Se$_3$ family  presents an AFM order while \textit{H}-CrSe$_2$ is non-magnetic (Fig. \ref{ES_mag}(c)). The rest of the Cr$_x$Se$_{x+1}$ series consists of magnetic metals, as seen in Fig. \ref{ES_mag}(b-c).

\textit{Cr$_2$Se$_3$}. The lowest-energy state of the bulk of Cr$_2$Se$_3$ corresponds to a FM ordering between out-of-plane first neighbours, and AF between in-plane atoms in Cr layers. At the GGA level, the electronic structure shows a band gap in the majority spin channel (Fig. S1(e)) while the minority spin channel shows a zero-gap band structure (Fig. S1(f)). At the 2D limit, for the 3-layer van der Waals unit with a vacancy, the favored magnetic order is FM both in-plane and out-of-plane. According to our DFT calculations, the magnetic moment of the Cr atoms is close to 3 $\mu_{B}$/Cr, which is the ionic value of a purely ionic Cr$^{3+}$ cation. In this case, the band structure shows an energy band gap that opens in both spin channels (Fig. S2(e-f)) in contrast to the bulk AF solution discussed earlier. 

Continuing to the 2D limit, in the case of the 2-layer structure, the out-of-plane magnetic configuration changes and the lowest-energy structure presents an AF order between Cr planes. This bilayer AF order reminds of the situation in CrI$_3$, also a Cr$^{3+}$ system where the off-plane coupling is AF in the bilayer\cite{huang2018electrical}. The insulating character of the system is also preserved (Fig. S3(e-f)), with spin up and down channels completely identical. The electronic structure of the system in this case can be well described based on the purely ionic picture where Cr$^{3+}$:d$^3$ cations are present in the compound, this being a very stable valence with a full majority t$_{2g}$ shell and a gap at the Fermi level. Of course, large covalency with the Se p bands exists, but the rough ionic picture survives and the system is a semiconductor. In this calculation, the magnetic moment of the Cr atoms according to DFT is again very close to 3 $\mu_{B}$/Cr, consistent with the nominal Cr$^{3+}$ valence close to the ionic limit. This structure represents the thinnest Cr$_2$Se$_3$ unit which preserves both stoichiometry and an octahedral environment formed by the Se atoms surrounding the Cr ions. 
Very recent measurements indicate this structure has a very strong FM in-plane coupling that leads to an enhanced magnetic transition temperature up to 230 K\cite{luout}.

The last of the structures for this same stoichiometry that we analyzed was a van der Waals bonded bilayer. The electronic structure suggests that the insulating nature of the compound is not affected by an increase in the number of van der Waals bonded layers (Fig. S4(e-f)). Again, the out-of-plane coupling is AF here and the band structure presents a t$_{2g}$-e$_g$ gap mainly dominated by states coming from nominally Cr$^{3+}$:d$^3$ cations.

\textit{CrSe$_2$}. Our calculations for the bulk 1\textit{T}-CrSe$_2$ system show that this bulk phase is a metal (Fig. S5(e-f)). It presents a magnetic moment around 2.15 $\mu_B$/Cr, consistent with the nominal Cr$^{4+}$:d$^2$ valence with substantial hybridizations. Also, monolayer 1\textit{T}-CrSe$_2$ (NS) has been \textit{ab initio} predicted to present a strain-induced switch from a FM phase to an AFM phase \cite{PhysRevB.92.214419}. Our calculations yield that bilayer CrSe$_2$, the coupling between planes is FM (Fig. S6(e-f)), different from previously published results indicating an in-plane FM and out-of-plane AFM coupling in the bulk \cite{PhysRevB.87.014420}. In terms of electronic structure, the monolayer is also a magnetic metal (Fig. S7(e-f)), similar to its bulk and bilayer counterparts. It has a magnetic moment around 2.2 $\mu_B$/Cr, again close to the ionic value for the nominally Cr$^{4+}$:d$^2$ cations, showing the existence of substantial Cr-Se hybridizations in the metallic limit. 

Our calculations also show that in the monolayer limit, FM coupling is the lowest-energy solution with a coupling strength $J=19$ meV, considering only first-neighbour interactions. Other published DFT calculations have also indicated that monolayer CrX$_2$ is an insulator whose bandgap decreases with increasing tensile strain \cite{doi:10.1021/jp501734s}. This is consistent with a metallic state in CrSe$_2$ under tensile strain that approaches the \textit{T} phase and a non-magnetic insulating behavior in bulk 2\textit{H}-CrSe$_2$. 
We have analyzed in more detail both phases, \textit{T} and \textit{H}, that show remarkable differences in their electronic structure. The 1\textit{T} structure has a magnetic metallic ground state with moments closer to high-spin Cr$^{4+}$:d$^2$, (with large covalency effects with Se \textit{p} bands). However, the 2\textit{H} structure shows a non-magnetic insulating state (Fig. S10(e-f)), presumably as a consequence of the strong covalency effects and the corresponding bonding-antibonding splittings that lead to a gap opening in the 2\textit{H} phase. 
When reducing the dimensionality of the bulk system down to an \textit{H}-bilayer, from inspection of the electronic structure of our calculations we see a similar energy band structure and DOS compared to the bulk case (Fig. S11(e-f)). In the monolayer limit, the electronic structure of the system (Fig. S12(e-f)) does not differ much from the bulk and bilayer cases. This indicates that the gap opening is not a quantum-confinement, layer-dependent effect, but instead it should be explained by covalent and crystal-field effects of the single layers. 

Looking at the electronic structure from a purely ionic crystal-field point of view, the observations are consistent. Since the lower-lying \textit{d}-orbitals in the trigonal prismatic crystal-field are the in-plane \textit{d}$_{x^2-y^2}$ and \textit{d}$_{xy}$ and these are degenerate, the only possible non-magnetic insulator is produced by a covalent combination of bonding-antibonding orbitals that opens a covalent gap at the Fermi level (otherwise, 2 electrons with opposite spins in 2 degenerate orbitals will always be a metal).
Because of these orbitals lying on the plane, they are basically unaffected by the out-of-plane configuration of the system, whether it is van der Waals bonded or not. The only orbitals affected by a dimensionality reduction would be those with a z component (out of plane). That is why the basic electronic structure around the Fermi level is unchanged by the dimensionality reduction.

\textit{NiAs-type structures}. Last, we analyze the stoichiometries which appear from reducing the dimensionality of the CrSe bulk system, (Fig. S15(f-g)). This  presents a magnetic moment of 3.9 $\mu_B$/Cr, very close to the high-spin ionic value for a Cr$^{2+}$:d$^4$ cation. The electronic structure of all the compounds of this NiAs-type family is very similar for L$\geq$3. From inspection of the electronic structure from our DFT calculations, both Cr$_3$Se$_4$ (3 Cr-layers) and Cr$_4$Se$_5$ (4 Cr-layers) are magnetic metals as shown in Fig. \ref{ES_mag}(b-c) (and in more detail in Fig. S13(f-g) and Fig. S14(f-g)). 
The electronic structure of the 3- and 4-layered systems differs greatly from that of the 2-layer Cr$_2$Se$_3$ system, which is a magnetic insulator. The magnetic moments of the 3- and 4-layer systems are 3.22 $\mu_B$/Cr and 3.37 $\mu_B$/Cr, respectively, reducing the e$_g$-band occupation that leads to the metallic behavior of these systems.

\section{Charge Density Waves}\label{sec:cdw}

\begin{figure*}
  \includegraphics[width=\textwidth]%
    {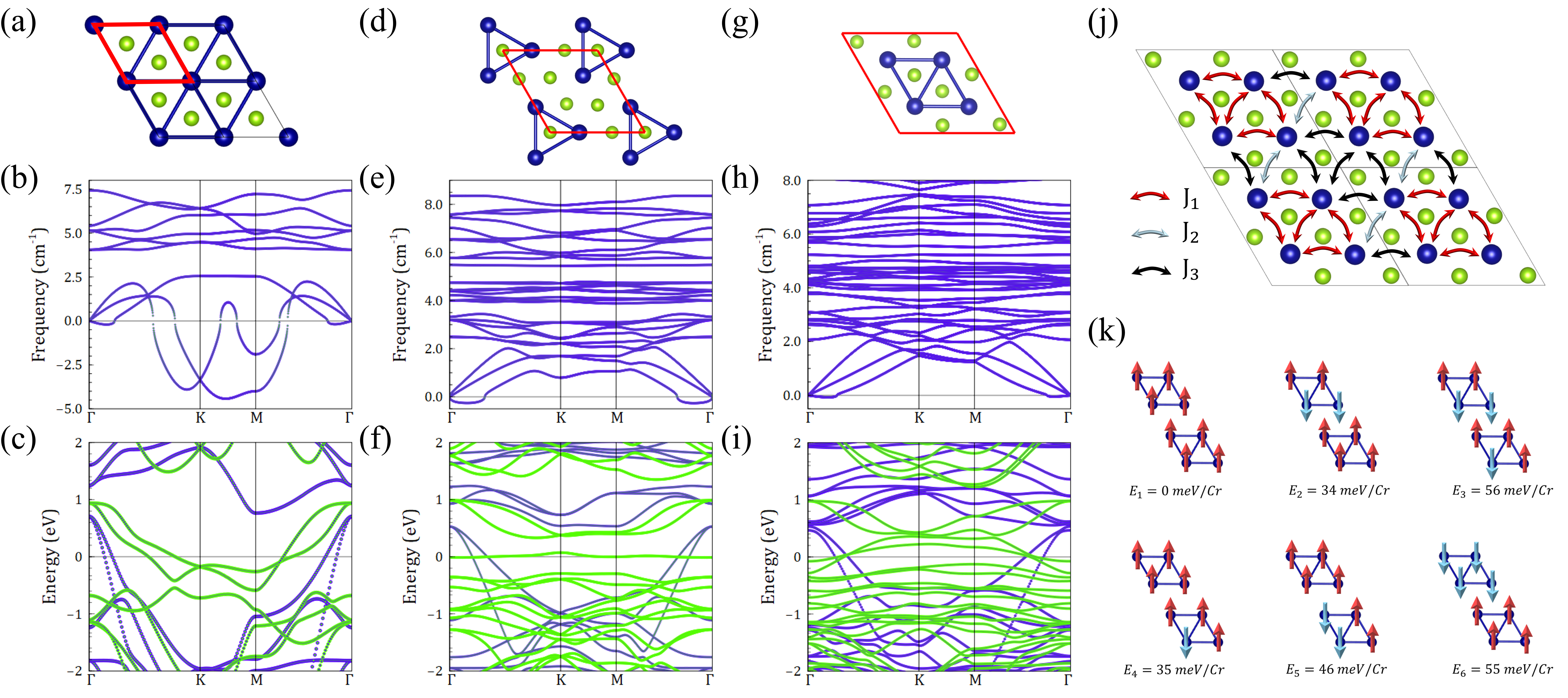}
     \caption{CrSe$_2$ CDW structures. (a) NS hexagonal high symmetry structure, its phonon band structure (b) and electronic band structure (c) where the majority spin channel is represented in green and the minority in blue. (d) CDW associated to the $\sqrt{3}\times\sqrt{3}$ supercell, its  phonon band structure (e) and electronic band structure (f) where the majority spin channel is represented in green and the minority in blue. (g) CDW associated to the $2\times2$ supercell, its  phonon band structure (h) and electronic band structure (i) where the majority spin channel is represented in green and the minority in blue. (j) Exchange couplings for the spin model. Double headed arrows indicate neighbour exchange interactions: Red arrows for first nearest neighbours ($J_1$), blue arrows for second nearest neighbours ($J_2$) and black arrows for third nearest neighbours ($J_3$). (k) Magnetic configurations used for the calculations of exchange couplings.}\label{cdw}
\end{figure*}

A ground state CDW phase will be associated to instabilities at certain \textit{q}-points related to the NS supercell, Fig. \ref{cdw}(a). In principle, these distortions are difficult to predict since many different orders may compete with each other. Previous studies report the emergence of different CDW phases in many TMD's \cite{eaglesham1986charge,fumega2019absence,mcmillan1975landau,rossnagel2011origin,wilson1974charge}, and these become more complicated when approaching the 2D limit \cite{coelho2019charge,diego2021van,fumega2022anharmonicity,otero2020controlled,xi2015strongly,lian2018unveiling,weber2011extended,bianco2019quantum,ugeda2016characterization}. The different unstable modes show a path towards possible reconstructions that can lead to a stable structure, Fig. \ref{cdw}(b). From the phonon band structure presented in Fig. \ref{cdw}(b) for 1\textit{T}-CrSe$_2$, the main instabilities are centered around \textit{K} and \textit{M}. The former would suggest an atomic reconstruction in a $\sqrt{3}\times\sqrt{3}$ supercell, while the latter would yield a $2\times2$ reconstruction. 

\textit{Supercell reconstructions in 1T-CrSe$_2$.} We start by analyzing the instability close to the high symmetry \textit{K} point ($\frac{1}{3}\ \frac{1}{3}\ $ 0), indicating a CDW phase associated to a $\sqrt{3}\times\sqrt{3}$ supercell. When performing the structural relaxation, the atoms in the monolayer relax into different positions to those of the high symmetry 1\textit{T}-phase. The new relaxed structure has an in-plane lattice parameter $a=5.99$ \r{A}, Fig. \ref{cdw}(d), close to the $\sqrt{3}\times\sqrt{3}$ lattice parameter with the $1\times1$ NS cell, $a=3.45$ \r{A}. This CDW is 32 meV/Cr lower in energy than the NS becoming a potential ground state for 1\textit{T}-CrSe$_2$. In addition, this superstructure presents stable phonon modes, shown in Fig. \ref{cdw}(e), despite the small imaginary frequencies that appear close to $\Gamma$, which are due to interpolation issues that occur in many 2D materials. Essentially, this lower-energy structure now consists of clusters of 3 Cr atoms. The bond distances shorten between the Cr forming the cluster, Fig. \ref{cdw}(d), contributing to the formation of a flat band (Fig. \ref{cdw}(f)) in the majority spin channel just at the Fermi level \cite{otero2020controlled,duan2024observation}. The magnetic moment is very similar to the monolayer, around 2.1 $\mu_B$/Cr, close to the ionic value of a Cr$^{4+}$:d$^2$ cation (substantially hybridized with Se).

The significant results obtained from the study of this $\sqrt{3}\times\sqrt{3}$ CDW have introduced a lower-energy state which is dynamically stable and presents a flat band at the Fermi level, which is an interesting platform for strong correlation effects. Figure \ref{cdw}(b) shows  another main instability at the M high symmetry point, corresponding to the \textit{q}-point (0.5, 0, 0) in the phonon band structure of the NS. We built a $2\times2$ supercell and relaxed the atomic positions, leading us to a new CDW, Fig. \ref{cdw}(g-h), with a lattice parameter of $a=6.90$ \r{A}, again very close to the monolayer NS value and identical to the $\sqrt{3}\times\sqrt{3}$ CDW. The new atomic configuration is lower in energy than the NS by $\sim$40 meV/Cr atom. By energy comparison, this $2\times2$ CDW is lower in energy than the $\sqrt{3}\times\sqrt{3}$ CDW by $\sim$ 8 meV/Cr atom, making it a probable ground state of the monolayer CrSe$_2$. The electronic structure describes a magnetic metal, with  magnetic moment of 2.1 $\mu_B$/Cr. The electronic band structure does not present a flat band at the Fermi level, Fig. \ref{cdw}(i). The $2\times2$ CDW reorganizes the  bands around the Fermi level leading to a further energy reduction (a high density of states near the Fermi level is usually an unstable situation that the system will try to avoid). Still, it is common that different CDW's compete in energy in TMD's, evolving with temperature in a non-trivial manner \cite{fumega2023anharmonicity}.

\textit{Magnetic couplings in 1T-CrSe$_2$.} Because of the possibility of this CDW in the $2\times2$ supercell being a ground state of the 1\textit{T} system, we went on to see how the magnetic couplings between the Cr atoms evolve. We have obtained the exchange interactions by fitting the total energy differences between various magnetic configurations to a Heisenberg model. The model was developed up to third Cr neighbours, as depicted in Fig. \ref{cdw}(j), with two unit cells containing 4 Cr atoms and 8 Se atoms per cell, and computed the energy differences between them, Fig. \ref{cdw}(k), that stand for the magnetic exchange couplings summarized on Table \ref{tab:CrSe2_mag_config} and Table \ref{tab:CrSe2_Js}. 

\setlength{\tabcolsep}{0.8em}
\begin{table}[h!]
    \centering
    \caption{Energies of the diferent magnetic configurations, using E1, the ferromagnetic state, as a reference. Results corresponding to the relaxed CrSe$_2$ lattice parameter.}
    \label{tab:CrSe2_mag_config}
    \begin{tabular}{|c|c|} \hline
          Mag. configurations & $\Delta$E (meV/Cr)  \\ \hline
         $E_1$  & 0.0  \\ \hline
         $E_2$  & 34 \\ \hline
         $E_3$  & 56  \\ \hline
         $E_4$  & 35  \\ \hline
         $E_5$  & 46  \\ \hline
         $E_6$  & 55  \\ \hline
    \end{tabular}
\end{table}

\setlength{\tabcolsep}{0.8em}
\begin{table}[h!]
    \caption{Exchange constants (in meV/Cr) for the different magnetic configurations used to calculate them. Results corresponding to the relaxed CrSe$_2$ lattice parameter. Each result corresponds to the indicated combination of magnetic configurations at the top of the column.}
    \label{tab:CrSe2_Js}
    \begin{tabular}{|c|c|c|c|c|} \hline
           & $E_1$, $E_3$, & $E_1$, $E_2$,   & $E_1$, $E_4$,   & $E_1$, $E_2$, \\  
           &  $E_5$, $E_6$ &  $E_3$, $E_6$ &   $E_5$, $E_6$ &   $E_4$, $E_5$ \\ \hline
         $J_1$  & 0.17   & 0.17   & 0.82  &  2.67 \\ \hline
         $J_2$  & -4.01  & 11.0  & -9.81 &  -0.39 \\ \hline
         $J_3$  & 17.7  & 12.7  & 19.7 &  10.3 \\ \hline
    \end{tabular}
\end{table}

These exchange interactions for the first and third neighbour exchange interactions (\textit{J}$_1$ and \textit{J}$_3$) are FM, but the second neighbour interactions (\textit{J}$_2$), oscillates between a FM and AFM depending on the configurations selected to do the mapping. Of course, this implies the limitations of using a Heisenberg model to describe a partly itinerant system where the moments are not completely localized. In any case, an AFM value of \textit{J}$_2$ would imply the existence of magnetic frustration in the system. 
We have also studied, using the same procedure, the behaviour of the exchange interactions under tensile strain. E.g., this could be realized by growing CrSe$_2$ on top of Cr$_2$Se$_3$ (that has a larger in-plane lattice parameter). 
The results, presented in Table \ref{tab:CrSe2_mag_config_strain} and Table \ref{tab:CrSe2_Js_strain}, show that the system responds reinforcing the FM nature of the exchange interactions compared to the values obtained for the relaxed lattice parameter of the CDW. No ambiguity occurs in this case, all the J's obtained from the different configurations are consistent, all of them are FM, and there is no trace of possible in-plane magnetic frustration in the system when tensile strain is introduced by using the Cr$_2$Se$_3$ lattice parameter for the calculations.

\setlength{\tabcolsep}{0.8em}
\begin{table}[h!]
    \caption{Energies of the different magnetic configurations, using E1, the ferromagnetic state, as a reference. Results corresponding to the relaxed Cr$_2$Se$_3$ lattice parameter.}
    \label{tab:CrSe2_mag_config_strain}
    \begin{tabular}{|c|c|} \hline
          Mag. configurations & $\Delta$E (meV/Cr)  \\ \hline
         $E_1$  & 0.0  \\ \hline
         $E_2$  & 83  \\ \hline
         $E_3$  & 132  \\ \hline
         $E_4$  & 58  \\ \hline
         $E_5$  & 99  \\ \hline
         $E_6$  & 98  \\ \hline
    \end{tabular}
\end{table}

\setlength{\tabcolsep}{0.8em}
\begin{table}[h!]
    \caption{Exchange constants (in meV/Cr) for the different magnetic configurations used to calculate them. Results corresponding to the relaxed Cr$_2$Se$_3$ lattice parameter. Each result corresponds to the indicated combination of magnetic configurations at the top of the column.}
    \label{tab:CrSe2_Js_strain}
    \begin{tabular}{|c|c|c|c|c|} \hline
           & $E_1$, $E_3$, & $E_1$, $E_2$,  & $E_1$, $E_4$,   & $E_1$, $E_2$, \\  
           & $E_5$, $E_6$ &   $E_3$, $E_6$ &  $E_5$, $E_6$ &  $E_4$, $E_5$ \\ \hline
         $J_1$  & 10 & 10  & 9.5  &  15 \\ \hline
         $J_2$  & 4.4  & 29  & 12 &  22 \\ \hline
         $J_3$  & 27 & 19  & 25 &  15 \\ \hline
    \end{tabular}
\end{table}


\section{Magnetocrystalline Anisotropy}\label{sec:MAE}

\begin{figure*}
  \includegraphics[width=\textwidth]
    {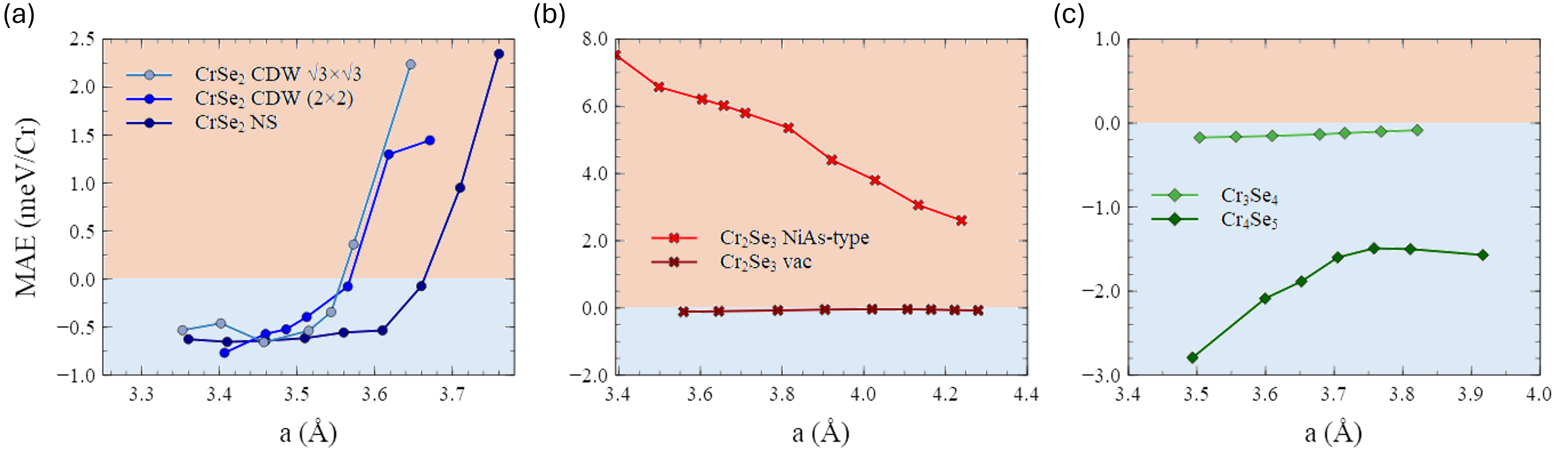}
     \caption{MAE as a function of in-plane strain. (a) MAE for the 1\textit{T}-CrSe$_2$ system, in particular, for the NS and the $\sqrt{3}$ $\times$ $\sqrt{3}$ and 2$\times$2 CDW reconstructions. MAE as a function of in-plane strain for the two different vdw Cr$_2$Se$_3$ units (b) and for Cr$_3$Se$_4$ and Cr$_4$Se$_5$ (c).}\label{mae_strain}
\end{figure*}

In order to unravel whether a long-range FM state could be sustained in 1\textit{T} CrSe$_2$ at the single-layer two-dimensional limit, we have computed the magnetocrystalline anisotropy energy. We have defined the magnetocrystalline anisotropy energy (MAE) so that a positive value indicates moments point out of the plane and a negative one means the in-plane solution is lower in energy. In the case that the moments prefer to point out of the plane, a long-range magnetic order could appear in the system at the ultra-thin limit.\cite{blei2021synthesis} The results are shown in Fig. \ref{mae_strain}(a) as a function of strain. Our results indicate that a finite-temperature FM order cannot occur for a wide range of in-plane lattice parameters around the relaxed value. Large amounts of strain are required to induce an out-of-plane anisotropy. However, the CDW's (both the 2 $\times$ 2 and the $\sqrt{3}$ $\times$ $\sqrt{3}$) lead to the stabilization of the moments in the out-of-plane configuration at smaller strains, as previously reported for CrTe$_2$.\cite{otero2020controlled}

When studying the magnetic anisotropy in systems which are not purely two-dimensional, the trend seems to be slightly different. In the case of Cr$_2$Se$_3$, Fig. \ref{mae_strain} (b), strain does not appear to flip the direction of the magnetic moments. For the case of the bilayer, where no Cr vacancies are present, the magnetic anisotropy points out-of-plane. Strain reduces the value of the MAE, but it does not reach a point where the direction of the magnetic moments flip. On the other hand, a structure which does present Cr vacancies has a MAE very close to zero, demonstrating a key role of vacancies in Cr$_2$Se$_3$ \cite{he2021two,li2023spontaneous}.

The magnetic anisotropy picture changes when analyzing systems with a number of layers higher than 2 in the NiAs-type structures i.e, increasing the number of layers in Cr$_2$Se$_3$. When analyzing the 3-layer case, i.e., Cr$_3$Se$_4$, the MAE has a small negative value, Fig. \ref{mae_strain}(c), inferring that the system is close to presenting no preferential magnetic direction. 
When adding another layer to the system, the MAE increases to a more negative value, as presented in Fig. \ref{mae_strain}(c). Thus, the in-plane direction will be the favoured orientation for the magnetic moments, demonstrating a strong tuning of the MAE as a function of the number of layers for NiAs-type structures without vacancies. 
On the other hand, we observe that strain does not appear do be an effective mechanism to control the preferential direction of the magnetic moments, Fig. \ref{mae_strain}(c), whereas control of local structure and stoichiometry does emerge as a key ingredient for quasi-2D magnets to emerge. This contrasts with the situation of the 1\textit{T}-CrSe$_2$ monolayer, where strain is a key ingredient in switching the MAE.

\section{Conclusions}\label{sec:conclusions}



In this paper we have adventured into the wide possibilities that the chromium selenide system offers, focusing on the transition from the bulk structure of CrSe$_2$ towards the single-layer limit.

A wide range of stoichiometries have been reported and with very different physical properties, and our goal is to provide a comprehensive first-principles analysis of them all. We began studying a known bulk structure, which is Cr$_2$Se$_3$ in its bulk form. This would be a nominally Cr$^{3+}$ system, with the Cr-t$_{2g}$ majority orbitals full, and an insulating gap at the Fermi level. From this bulk structure, we tried to isolate a structure which was as close as possible to the 2D limit. We managed to computationally engineer a 3-layer structure, with vacancies, imitating the bulk structure. We also managed to build a thinner layered structure for this stoichiometry. We computationally confirmed that a 2-layer system, that constitutes a smaller unit building block, may grow in a van der Waals bonded layered structure. Both these structures are magnetic insulators, and the main difference we observed to tell them apart was the MAE. The 3-layered vacancy structure does not present an in-plane or out-of-plane favoured direction for the magnetic moments to point at, but the 2-layer system does. Its magnetic moments point out-of-plane and its magnetic anisotropy is very robust to strain. The value of its MAE decreases when applying strain but does not appear to flip the favoured magnetization direction. This means that we can differentiate for this thin layered Cr$_2$Se$_3$ structure whether it presents vacancies or not. If it does, there will be no long-range FM order, but if it does not, one could have it at a finite temperature. The in-plane top view of the structures are identical, and the lattice parameter of both 2D-like structures lies close to $3.65$ \r{A}. This means that the analysis of the MAE is the main tool to identify experimentally which structure one is working with. This is a system that can be driven to the 2D limit retaining the basic structure and stoichiometry.

We then went to analyze a more 2D approachable stoichiometry by studying CrSe$_2$. We started studying the bulk structure, observing that it is a magnetic metal that is not dynamically stable. Reducing the dimensionality of the system down to a bilayer did not change the basic picture. The in-plane imaginary phonon modes still exist, and the electronic structure does not present any major differences compared to the bulk. The next step was studying the actual monolayer, which again does not differ much from the bilayer and bulk structure in its electronic structure or dynamic stability. Based on the instability at the K \textit{q}-point we analyzed a possible CDW in a $\sqrt{3}\times\sqrt{3}$ supercell, and found it a lower-energy state that is dynamically stable. We observed that the electronic structure of this system changed compared to the NS. A flat band appears at the Fermi level in the majority spin channel. This is of great interest because it is a platform for strong correlations to dominate. Even if it is a metastable solution, many of these CDW metastable phases can be eventually synthesized at the appropriate growth conditions\cite{di2023metastable}. Taking into account that there was another major instability around the M \textit{q}-point, we went on and repeated the previous procedure but for a $2\times2$ supercell. Again, we found a CDW that is a dynamically stable lower-energy state, $\sim$8 meV/Cr atom lower in energy compared to the $\sqrt{3}\times\sqrt{3}$ CDW, meaning that this new CDW is a possible ground state of the monolayer CrSe$_2$ system. This CDW maintains the common trait of all of the 1\textit{T}-CrSe$_2$ systems that we analyzed, as it is also a magnetic metal. All of the systems studied present reduced magnetic moments  compared with the nominal ionic Cr$^{3+}$ value. The in-plane lattice parameter of all of this \textit{T}-phase family are very close, all around $3.45-3.46$ \r{A}. This makes them distinguishable from the different Cr$_2$Se$_3$ structures, but very difficult to distinguish experimentally from the 1\textit{T}-CrSe$_2$ family by their lattice parameter. In order to tell them apart, STM can be a good solution, as it would show clear structural differences due to the CDW's that are present. The analysis of the MAE of the monolayers shows that the appearance of any of the two CDW favours magnetic long-range order in the system, because the strain required is reduced compared to what is needed in the NS. This means that the CDW's in CrSe$_2$ are a key ingredient in order to engineer 2D magnets in this system.
The next step is to study the \textit{H}-phase of the CrSe$_2$ stoichiometry. This was motivated by completeness of our analysis and results from the literature which indicated that this phase is a lower-energy state compared to the 1\textit{T}-phase family. The starting point was studying the bulk, which presents a dynamically stable structure, unlike the 1\textit{T} phase. In this case, the system is a non-magnetic insulator. These properties do not change when reducing the dimensionality down to the bilayer or the monolayer. The lattice parameter of the \textit{H}-phase family is around $3.20$ \r{A}, which makes it easily distinguishable from the different Cr$_2$Se$_3$ structures and from its 1\textit{T} counterpart. The difference in the Se atom distribution would also be an interesting way of telling these two phases apart.

Our final analysis lies in the possibility of different stoichiometries constructed by building up/down non van der Waals bonded layers from the CrSe NiAs-type structure. This structure is present in the Cr$_x$Se$_{x+1}$ family, as can be seen in the bulk structure of Cr$_2$Se$_3$. Its bulk structure is basically the NiAs one but with Cr-vacancies in alternating layers. We observed that, if no vacancies were present in the system, one could relate the stoichiometry of these systems with the number of layers L in such a way that Cr$_L$Se$_{L+1}$. So, when $L=1$ we would recover the 1\textit{T}-phase analysis, and for $L=2$ we would recover the 2 Cr-layered monolayer that we presented for Cr$_2$Se$_3$. We analyzed the $L=3, 4$ cases, as well as the infinite-layer situation, which leads us back to CrSe in its bulk structure. All of these new structures present progressively increasing lattice parameters. The $L=3$ layered structure which would correspond to Cr$_3$Se$_4$ presents an in-plane lattice parameter of $a=3.67$ \r{A}, for $L=4$ we have Cr$_4$Se$_5$ with an in-plane lattice parameter of $a=3.70$ \r{A}, and finally the infinite-layer structure of CrSe bulk has a lattice parameter of $a=3.76$ \r{A}. As we can see, the in-plane lattice parameter increases with the number of layers. All of these stoichiometries present similar electronic structures, as they are magnetic metals and their MAE has the magnetic moments pointing in the in-plane direction. None of them presents any imaginary phonon modes, so they are all dynamically stable. This makes them excellent candidates for possible structures in different Cr$_x$Se$_{x+1}$ van der Waals layered stoichiometries.

\section*{Acknowledgements}
This work is supported by the MINECO of Spain through the projects PGC2018-101334-B-C21,  PGC2018-101334-A-C22, PID2021-122609NB-C21 and PID2021-122609NB-C22. We thank the CESGA (Centro de Supercomputacion de Galicia) for the computing facilities provided. J.P. thanks MECD for the financial support received through the ``Ayudas para contratos predoctorales para la formación de doctores" grant PRE2019-087338. A. O. F. thanks the financial support from the Academy of Finland Project No. 349696.

\bibliography{refs}

\end{document}


\title{Supplemental Material: An ab initio description of the family of Cr selenides: structure, magnetism and electronic structure from bulk to the single-layer limit}

\author{Jan Phillips}
  \email{j.phillips@usc.es}
\affiliation{Departamento de F\'{i}sica Aplicada,
  Universidade de Santiago de Compostela, E-15782 Campus Sur s/n,
  Santiago de Compostela, Spain}
\affiliation{Instituto de Materiais iMATUS, Universidade de Santiago de Compostela, E-15782 Campus Sur s/n, Santiago de Compostela, Spain}   
\author{Adolfo O. Fumega}
\affiliation{Department of Applied Physics, Aalto University, 02150 Espoo, Finland}

\author{S. Blanco-Canosa}
\affiliation{Donostia International Physics Center (DIPC), San Sebastián, Spain}
\affiliation{IKERBASQUE, Basque Foundation for Science, 48013 Bilbao, Spain}

\author{Victor Pardo}
  \email{victor.pardo@usc.es}
\affiliation{Departamento de F\'{i}sica Aplicada,
  Universidade de Santiago de Compostela, E-15782 Campus Sur s/n,
  Santiago de Compostela, Spain}
\affiliation{Instituto de Materiais iMATUS, Universidade de Santiago de Compostela, E-15782 Campus Sur s/n, Santiago de Compostela, Spain}

\maketitle

\section{Computational Details}
Here we provide additional details of all the density functional theory (DFT) calculations presented in the main text and below.

In particular, when computing with {\sc wien2k}, all results where obtained using a value of R$_{mt}$K$_{max}$=7.0. 
The R$_{mt}$ values used were the same for the Cr and the Se atoms. Depending on the compounds, the values in a.u. used were the following: 2.29 for the bulk and trilayer structure of Cr$_2$Se$_3$, 2.10 for the bilayer structure of Cr$_2$Se$_3$, 2.13 for the NS \textit{T} and \textit{H} monolayer and bilayer phases of CrSe$_2$, 2.00 for the $2\times2$ and the $\sqrt{3}\times\sqrt{3}$ CDW phases of \textit{T} CrSe$_2$, 2.15 for the NS \textit{T} and \textit{H} bulk phases of CrSe$_2$, 2.19 for the Cr$_3$Se$_4$ and Cr$_3$Se$_4$, and 2.24 for CrSe bulk.
When computing with {\sc VASP}, all results where obtained with the following cutoffs: The cut-off energy of the plane wave representation of the augmentation charges was ENAUG = 450 eV and the cutoff energy for the plane-wave-basis set was ENCUT = 230 eV.

\section{Analyzed compounds}

In order to obtain our results we have analyzed a series of different stoichiometries from the Cr$_x$Se$_y$ family and different structures presented within the same stoichiometry. In this section we present a fingerprint of each one, consisting of a top, side and slanted view of the structure, electronic band structure for the majority (green) and minority (blue) spin channels, and phonon band structure of each compound discussed in the main text. 

\subsection{Cr$_2$Se$_3$}

In Fig. \ref{Cr2Se3_bulk_1T_NS} we present the fingerprint for bulk Cr$_2$Se$_3$. The structure consists of alternating Cr layers with no vacancies with layers that have 2 vacancies per unit cell. 

\begin{figure}[!h]
  \centering
  \includegraphics[width=0.9\textwidth]%
    {figures/figures_supplementary/Fig_bulk_Cr2Se3_1T_NS.png}
     \caption{Structure of bulk Cr$_2$Se$_3$. (a) Top view, where the triangular coordination of Cr atoms forms a hexagon and the red rhombus represents the unit cell. Side view (b), and slanted view (c), where the octahedral environment formed by the Se atoms surrounding the Cr atoms can be seen, as well as the vacuum needed to form the monolayer. (d) Phonon band structure computed for the 2$\times$2$\times$1 supercell. Electronic structure for both the majority (e) and minority (f) spin channels where the band structure and DOS of both channels are presented.}\label{Cr2Se3_bulk_1T_NS}
\end{figure}

In Fig. \ref{Cr2Se3_3layer_1T_NS} we present the fingerprint for a 3-layer possible van der Waals Cr$_2$Se$_3$ unit. We engineered this structure in order to preserve the stoichiometry when approaching the 2D limit. The structure presents two Cr-rich layers with a Cr-vacancy layer sandwiched in between them. There is only one vacancy per unit cell in this structure.

\begin{figure}[!h]
  \centering
  \includegraphics[width=\textwidth]%
    {figures/figures_supplementary/Fig_3layer_Cr2Se3_1T_NS.png}
     \caption{Structure of monolayer 1\textit{T} Cr$_2$Se$_3$ with one Cr-vacancy layer. (a) Top view, where the triangular coordination of Cr atoms forms a hexagon and the red rhombus represents the unit cell. Side view (b), and slanted view (c), where the 1\textit{T} octahedral environment formed by the Se atoms surrounding the Cr atoms can be seen, as well as the vacuum needed to form the monolayer together with no vdW gap between layers. (d) Phonon band structure computed for the 2$\times$2$\times$1 (left) and the 3$\times$3$\times$1 (right) supercells. Electronic structure for both the majority (e) and minority (f) spin channels where the band structure and DOS of both channels are presented.}\label{Cr2Se3_3layer_1T_NS}
\end{figure}

In Fig. \ref{Cr2Se3_1layer_1T_NS} we present the fingerprint for a two layered possible van der Waals Cr$_2$Se$_3$ unit. We engineered this structure in order to preserve the stoichiometry when approaching the 2D limit. The structure presents two Cr-rich layers with no vacancies in the unit cell.

\begin{figure}[!h]
  \centering
  \includegraphics[width=\textwidth]%
    {figures/figures_supplementary/Fig_1layer_Cr2Se3_1T_NS.png}
     \caption{Structure of monolayer 1\textit{T} Cr$_2$Se$_3$ originated from 2 Cr-rich layers. (a) Top view, where the triangular coordination of Cr atoms forms a hexagon and the red rhombus represents the unit cell. Side view (b), and slanted view (c), where the 1\textit{T} octahedral environment formed by the Se atoms surrounding the Cr atoms can be seen, as well as the vacuum needed to form the monolayer together with no vdW gap between layers. (d) Phonon band structure computed for the 2$\times$2$\times$1 (left) and the 3$\times$3$\times$1 (right) supercells. Electronic structure for both the majority (e) and minority (f) spin channels where the band structure and DOS of both channels are presented.}\label{Cr2Se3_1layer_1T_NS}
\end{figure}

In Fig. \ref{Cr2Se3_2layer_1T_NS} we present the fingerprint for a bilayer of the previous two layered possible van der Waals Cr$_2$Se$_3$ unit. 

\begin{figure}[!h]
  \centering
  \includegraphics[width=\textwidth]%
    {figures/figures_supplementary/Fig_2layer_Cr2Se3_1T_NS.png}
     \caption{Structure of bilayer 1\textit{T} Cr$_2$Se$_3$. Engineered from two monolayers of 1\textit{T} Cr$_2$Se$_3$ originated from 2 Cr-rich layers. (a) Top view, where the triangular coordination of Cr atoms forms a hexagon and the red rhombus represents the unit cell. Side view (b), and slanted view (c), where the 1\textit{T} octahedral environment formed by the Se atoms surrounding the Cr atoms can be seen, as well las the vaccuum neded to form the bilayer together with the vdW gap between layers. (d) Phonon band structure computed for the 2$\times$2$\times$1 (left) and the 3$\times$3$\times$1 (right) supercells. Electronic structure for both the majority (e) and minority (f) spin channels where the band structure and DOS of both channels are presented.}\label{Cr2Se3_2layer_1T_NS}
\end{figure}

\subsection{CrSe$_2$}

In Fig. \ref{CrSe2_bulk_1T_NS} we present the fingerprint for bulk CrSe$_2$. The structure consists of van der Waals bonded 1\textit{T}-CrSe$_2$ layers.

\begin{figure}[!h]
  \centering
  \includegraphics[width=\textwidth]%
    {figures/figures_supplementary/Fig_bulk_CrSe2_1T_NS.png}
     \caption{Structure of bulk 1\textit{T} CrSe$_2$. (a) Top view, where the triangular coordination of Cr atoms forms a hexagon and the red rhombus represents the unit cell. Side view (b), and slanted view (c), where the octahedral environment formed by the Se atoms surrounding the Cr atoms can be seen, as well as the vdW gap between layers. (d) Phonon band structure computed for the 6$\times$6$\times$2 supercell. Electronic structure for both the majority (e) and minority (f) spin hannels where the band structure and DOS of both channels are presented.}\label{CrSe2_bulk_1T_NS}
\end{figure}

In Fig. \ref{CrSe2_2layer_1T_NS} we present the fingerprint for bilayer 1\textit{T}-CrSe$_2$. The structure consists of two van der Waals bonded 1\textit{T}-CrSe$_2$ layers.

\begin{figure}[!h]
  \centering
  \includegraphics[width=\textwidth]%
    {figures/figures_supplementary/Fig_2layer_CrSe2_1T_NS.png}
     \caption{Structure of bilayer 1\textit{T} CrSe$_2$. (a) Top view, where the triangular coordination of Cr atoms forms a hexagon and the red rhombus represents the unit cell. Side view (b), and slanted view (c), where the octahedral environment formed by the Se atoms surrounding the Cr atoms can be seen, as well as the vacuum needed to form the bilayer together with the vdW gap between layers. (d) Phonon band structure computed for the 6$\times$6$\times$1 supercell. Electronic structure for both the majority (e) and minority (f) spin channels where the band structure and DOS of both channels are presented.}\label{CrSe2_2layer_1T_NS}
\end{figure}

In Fig. \ref{CrSe2_monolayer_1T_NS} we present the fingerprint for monolayer 1\textit{T}-CrSe$_2$. The structure consists of a single 1\textit{T}-CrSe$_2$ layer.

\begin{figure}[!h]
  \centering
  \includegraphics[width=\textwidth]%
    {figures/figures_supplementary/Fig_monolayer_CrSe2_1T_NS.png}
     \caption{Structure of monolayer 1\textit{T} CrSe$_2$. (a) Top view, where the triangular coordination of Cr atoms forms a hexagon and the red rhombus represents the unit cell. Side view (b), and slanted view (c), where the octahedral environment formed by the Se atoms surrounding the Cr atoms can be seen, as well as the vacuum needed to form the monolayer. (d) Phonon band structure computed for the 6$\times$6$\times$1 supercell. Electronic structure for both the majority (e) and minority (f) channels. Band structure and DOS of both channels are presented.}\label{CrSe2_monolayer_1T_NS}
\end{figure}

In Fig. \ref{CrSe2_monolayer_1T_CDW_r3xr3} we present the fingerprint for the monolayer CDW $\sqrt{3}\times\sqrt{3}$ reconstruction of the 1\textit{T} CrSe$_2$ structure. This reconstruction was obtained from the instability close to the K-point seen by inspection of the phonon band structure in Fig. \ref{CrSe2_monolayer_1T_NS}(d).

\begin{figure}[!h]
  \centering
  \includegraphics[width=\textwidth]%
    {figures/figures_supplementary/Fig_monolayer_CrSe2_1T_CDW_r3xr3.png}
     \caption{Structure of monolayer CDW $\sqrt{3}\times\sqrt{3}$ reconstruction of the 1\textit{T} CrSe$_2$ structure. (a) Top view, where the Cr atoms group into 3-atom triangular clusters of nearest neighbours, and the red rhombus represents the unit cell. Side view (b), and slanted view (c), where the octahedral environment formed by the Se atoms surrounding the Cr atoms can be seen, as well as the vacuum needed to form the monolayer. (d) Phonon band structure computed for the 2$\times$2$\times$1 (left) and 3$\times$3$\times$1 (right) supercells. Electronic structure for both the majority (e) and minority (f) spin channels where the band structure and DOS of both channels are presented.}\label{CrSe2_monolayer_1T_CDW_r3xr3}
\end{figure}

In Fig. \ref{CrSe2_monolayer_1T_CDW_2x2} we present the fingerprint for the monolayer CDW 2$\times$2 reconstruction of the 1\textit{T} CrSe$_2$ structure. This reconstruction was obtained from the instability close to the M-point seen by inspection of the phonon band structure in Fig. \ref{CrSe2_monolayer_1T_NS}(d)

\begin{figure}[!h]
  \centering
  \includegraphics[width=\textwidth]%
    {figures/figures_supplementary/Fig_monolayer_CrSe2_1T_CDW_2x2.png}
     \caption{Structure of monolayer CDW 2$\times$2 reconstruction of the 1\textit{T} CrSe$_2$ structure. Top view, where the Cr atoms group into 4-atom rhombohedral clusters of nearest neighbours, and the red rhombus represents the unit cell. Side view (b), and slanted view (c), where the octahedral environment formed by the Se atoms surrounding the Cr atoms can be seen, as well as the vacuum needed to form the monolayer. (d) Phonon band structure computed for the 2$\times$2$\times$1 supercell. Electronic structure for both the majority (e) and minority (f) spin channels where the band structure and DOS of both channels are presented.}\label{CrSe2_monolayer_1T_CDW_2x2}
\end{figure}

In Fig. \ref{CrSe2_bulk_2H_NS} we present the fingerprint for the bulk 2\textit{H}-CrSe$_2$ structure. The structure consists of van der Waals bonded \textit{H}-CrSe$_2$ layers.

\begin{figure}[!h]
  \centering
  \includegraphics[width=0.95\textwidth]%
    {figures/figures_supplementary/Fig_bulk_CrSe2_2H_NS.png}
     \caption{Structure of bulk 2\textit{H} CrSe$_2$. (a) Top view, where the triangular coordination of Cr atoms forms a hexagon and the red rhombus represents the unit cell. Side view (b), and slanted view (c), where the tetragonal environment formed by the Se atoms surrounding the Cr atoms can be seen,  as well as the vdW gap between layers. (d) Phonon band structure computed for the 2$\times$2$\times$2 (left) and the 3$\times$3$\times$2 (right) supercells. Electronic structure for both the majority (e) and minority (f) spin channels where the band structure and DOS of both channels are presented.}\label{CrSe2_bulk_2H_NS}
\end{figure} 

In Fig. \ref{CrSe2_2layer_2H_NS} we present the fingerprint for bilayer \textit{H}-CrSe$_2$ structure. The structure consists of two van der Waals bonded \textit{H}-CrSe$_2$ layers. 

\begin{figure}[!h]
  \centering
  \includegraphics[width=0.9\textwidth]%
    {figures/figures_supplementary/Fig_2layer_CrSe2_2H_NS.png}
     \caption{Structure of bilayer 2\textit{H} CrSe$_2$. (a) Top view, where the triangular coordination of Cr atoms forms a hexagon and the red rhombus represents the unit cell. Side view (b), and slanted view (c), where the tetragonal environment formed by the Se atoms surrounding the Cr atoms can be seen,  as well as the vacuum needed to form the bilayer together with the vdW gap between layers. (d) Phonon band structure computed for the 3$\times$3$\times$1 (left) and the 4$\times$4$\times$1 (right) supercells. Electronic structure for both the majority (e) and minority (f) spin channels where the band structure and DOS of both channels are presented.}\label{CrSe2_2layer_2H_NS}
\end{figure}

In Fig. \ref{CrSe2_monolayer_1H_NS} we present the fingerprint for monolayer \textit{H}-CrSe$_2$ structure. The structure consists of one \textit{H}-CrSe$_2$ layer.

\begin{figure}[!h]
  \centering
  \includegraphics[width=\textwidth]%
    {figures/figures_supplementary/Fig_monolayer_CrSe2_1H_NS.png}
     \caption{Structure of monolayer \textit{H} CrSe$_2$. (a) Top view, where the triangular coordination of Cr atoms forms a hexagon and the red rhombus represents the unit cell. Side view (b), and slanted view (c), where the tetragonal environment formed by the Se atoms surrounding the Cr atoms can be seen,  as well as the vacuum needed to form the monolayer. (d) Phonon band structure computed for the 6$\times$6$\times$1 supercell. Electronic structure for both the majority (e) and minority (f) spin channels where the band structure and DOS of both channels are presented.}\label{CrSe2_monolayer_1H_NS}
\end{figure}

\subsection{Other NiAs-type structures}

In Fig. \ref{Cr3Se4_1layer_1T_NS} we present the fingerprint for a possible van der Waals Cr$_3$Se$_4$ unit. We engineered this structure from a NiAs-type 3 layered structure, which corresponds to this stoichiometry.

\begin{figure}[!h]
  \centering
  \includegraphics[width=0.9\textwidth]%
    {figures/figures_supplementary/Fig_1layer_Cr3Se4_1T_NS.png}
     \caption{Structure of monolayer Cr$_3$Se$_4$ originated from 3 layers of the CrSe bulk NiAs-type structure. (a) Top view, where the triangular coordination of Cr atoms forms a hexagon and the red rhombus represents the unit cell. Side view (b), and slanted view (c), where the 1\textit{T} octahedral environment formed by the Se atoms surrounding the Cr atoms can be seen, as well as the vacuum needed to form the monolayer together with no vdW gap between layers. (d) Phonon band structure computed for the 2$\times$2$\times$1 (left) and the 3$\times$3$\times$1 (right) supercells. Electronic structure for both the majority (e) and minority (f) spin channels where the band structure and DOS of both channels are presented.}\label{Cr3Se4_1layer_1T_NS}
\end{figure}

In Fig. \ref{Cr4Se5_1layer_1T_NS} we present the fingerprint for a possible van der Waals Cr$_4$Se$_5$ unit. We engineered this structure from a NiAs-type 4 layered structure, which corresponds to this stoichiometry.

\begin{figure}[!h]
  \centering
  \includegraphics[width=0.9\textwidth]%
    {figures/figures_supplementary/Fig_1layer_Cr4Se5_1T_NS.png}
     \caption{Structure of monolayer Cr$_4$Se$_5$ originated from 4 layers of the CrSe bulk NiAs-type structure (a) Top view, where the triangular coordination of Cr atoms forms a hexagon and the red rhombus represents the unit cell. Side view (b), and slanted view (c), where the 1\textit{T} octahedral environment formed by the Se atoms surrounding the Cr atoms can be seen, as well as the vacuum needed to form the monolayer together with no vdW gap between layers. (d) Phonon band structure computed for the 2$\times$2$\times$1 (left) and the 3$\times$3$\times$1 (right) supercells. Electronic structure for both the majority (e) and minority (f) spin channels where the band structure and DOS of both channels are presented.}\label{Cr4Se5_1layer_1T_NS}
\end{figure}

In Fig. \ref{CrSe_bulk_1T_NS} we present the fingerprint for a possible bulk structure of CrSe. We engineered this structure from a NiAs-type bulk (or infinit layered) structure, which corresponds to this stoichiometry.

\begin{figure}[!h]
  \centering
  \includegraphics[width=\textwidth]%
    {figures/figures_supplementary/Fig_bulk_CrSe_1T_NS.png}
     \caption{Structure of bulk CrSe in the NiAs-type structure. (a) Top view, where the triangular coordination of Cr atoms forms a hexagon and the red rhombus represents the unit cell. Side view (b), and slanted view (c), where the 1\textit{T} octahedral environment formed by the Se atoms surrounding the Cr atoms can be seen, with no vdW gap between layers. (d) Phonon band structure computed for the 2$\times$2$\times$2 (left) and the 3$\times$3$\times$2 (right) supercells. Electronic structure for both the majority (e) and minority (f) spin channels where the band structure and DOS of both channels are presented.}\label{CrSe_bulk_1T_NS}
\end{figure}